\documentclass[twocolumn,pra,english,reprint, longbibliography, superscriptaddress, breaklinks=true, showkeys, showpacs=false, nofootinbib]{revtex4}

\usepackage[T1]{fontenc}
\usepackage[latin9]{inputenc}
\usepackage{color}
\usepackage{babel}
\usepackage{amsmath}
\usepackage{amssymb}
\usepackage{subfigure}
\usepackage{graphicx}
\usepackage{physics}
\usepackage{float}
\usepackage{lipsum}  
\usepackage{comment}
\usepackage[unicode=true,pdfusetitle,
 bookmarks=true,bookmarksnumbered=false,bookmarksopen=false,breaklinks=true,pdfborder={0 0 0},backref=false,colorlinks=true]
 {hyperref} 
\makeatletter
\@ifundefined{textcolor}{}
{%
 \definecolor{BLACK}{gray}{0}
 \definecolor{WHITE}{gray}{1}
 \definecolor{RED}{rgb}{1,0,0}
 \definecolor{GREEN}{rgb}{0,1,0}
 \definecolor{BLUE}{rgb}{0,0,1}
 \definecolor{CYAN}{cmyk}{1,0,0,0}
 \definecolor{MAGENTA}{cmyk}{0,1,0,0}
 \definecolor{YELLOW}{cmyk}{0,0,1,0}
}
\pdfoutput=1
\hypersetup{colorlinks=true,citecolor=blue,linkcolor=cyan,urlcolor=blue,filecolor= green, breaklinks=true}
\usepackage{url}
\usepackage{breakurl}
\makeatother

\begin{document}

\author{Vitor Dallabona}
\email{vitor.dallabona@acad.ufsm.br}
\affiliation{Department of Electronics and Computing, Technology Center, Federal University of Santa Maria, 97105-900, Santa Maria, RS, Brazil}

\author{Lucas Friedrich}
\email{lucas.friedrich@acad.ufsm.br}
\affiliation{Physics Department, Center for Natural and Exact Sciences, Federal University of Santa Maria, 97105-900, Santa Maria, RS, Brazil}

\author{Jonas Maziero}
\email{jonas.maziero@ufsm.br}
\affiliation{Physics Department, Center for Natural and Exact Sciences, Federal University of Santa Maria, 97105-900, Santa Maria, RS, Brazil}

\selectlanguage{english}%

\title{Resource Analysis of Quantum Optimization Algorithms via Variational Methods}

\begin{abstract}
Variational Quantum Algorithms (VQAs) are promising approaches for solving combinatorial optimization problems on near-term quantum devices. However, the role played by quantum resources such as entanglement and coherence during the optimization process remains poorly understood. In this work, we investigate the evolution of these resources in two VQAs--the Quantum Approximate Optimization Algorithm (QAOA) and the Feedback-based Algorithm for Quantum Optimization (FALQON)--applied to the Minimum Vertex Cover problem. We employ a variational resource-quantification framework based on the Bures metric and Uhlmann's theorem, in which an auxiliary quantum circuit is optimized to identify the closest resource-free states. This approach allows us to quantify Bures entanglement and Bures coherence of the algorithmic states without requiring full quantum state tomography. For the considered problem instance, QAOA exhibits a pronounced accumulation of entanglement in its initial layers, followed by a progressive reduction as the algorithm converges toward the optimal solution. Its coherence likewise decreases overall with increasing circuit depth, although a finite residual value persists due to the degeneracy of the optimal-solution subspace. In contrast, FALQON exhibits a substantially lower entanglement peak, reached at larger circuit depth, followed by a decay as convergence proceeds. Its coherence decreases smoothly and monotonically together with the optimization error. These results reveal distinct quantum-resource trajectories associated with the two optimization strategies and illustrate how entanglement and coherence evolve as variational quantum states approach solutions of a classical combinatorial optimization problem.
\end{abstract}


\date{\today}

\maketitle

\section{Introduction}


The Noisy Intermediate-Scale Quantum (NISQ) era represents the current stage of quantum computing, in which available devices operate with a limited number of qubits and are subject to significant hardware constraints. The absence of fault-tolerant quantum error correction (QEC) \cite{Guedes2026,Campbell2024}, together with noise and limited coherence times \cite{Preskill2018,Bharti2022}, degrades operational fidelity and limits the execution of deep quantum circuits and large-scale algorithms. These limitations motivate the development of approaches that can effectively exploit the capabilities of near-term quantum devices. In this context, Variational Quantum Algorithms (VQAs) \cite{Cerezo2021,Qi2024} have emerged as a promising paradigm, combining parameterized quantum circuits (PQCs) with classical optimization routines to tackle relevant computational problems under NISQ-era constraints.

VQAs exploit a hybrid quantum-classical architecture in which a parameterized quantum circuit is iteratively updated by a classical optimizer to minimize a cost function associated with the problem of interest. By combining quantum state preparation and measurement with classical optimization, this approach provides a practical framework for exploring computational problems on near-term quantum devices, although its performance remains constrained by noise, circuit depth, and optimization challenges \cite{HurT2022, Peruzzo2014}. In quantum chemistry, a prominent example is the Variational Quantum Eigensolver (VQE) \cite{Peruzzo2014,Tilly2022}, which estimates molecular ground-state energies by minimizing the expectation value of a molecular Hamiltonian. In combinatorial optimization, the Quantum Approximate Optimization Algorithm (QAOA) \cite{Farhi2014} and its variants \cite{Blekos2024} encode objective functions into cost Hamiltonians whose ground states represent optimal solutions \cite{Lucas2014}. However, the classical optimization of variational parameters can become challenging, motivating alternative approaches such as the Feedback-based Algorithm for Quantum Optimization (FALQON) \cite{Magann2022}. FALQON uses measurement-based feedback to determine circuit parameters through a prescribed control rule, thereby avoiding a conventional classical optimization loop. Building on this approach, FALQON+ \cite{Magann2022-1} uses parameters obtained from a shallow FALQON execution to initialize QAOA, combining feedback-based parameter selection with subsequent classical optimization in an effort to improve optimization efficiency and solution quality.


Understanding how quantum resources, particularly coherence and entanglement, influence the performance of VQAs is essential for assessing their behavior and developing strategies to improve their effectiveness on noisy, near-term quantum devices. However, quantifying these resources presents a significant challenge, as conventional approaches often rely on quantum state tomography \cite{James2001,Gross2010}, whose experimental and computational costs grow rapidly with system size or require solving additional optimization problems. These limitations motivate the development of resource-quantification methods that can characterize quantum states without requiring full state reconstruction.


In this work, we apply the quantum resource quantification method proposed in Ref. \cite{Friedrich2025} to investigate the evolution of quantum coherence and entanglement in the QAOA, FALQON, FALQON+, and VQE algorithms. Based on the Bures distance and Uhlmann's theorem \cite{Uhlmann1976, Jozsa1994}, this method formulates resource quantification as a variational optimization problem, enabling the estimation of these resources without requiring full quantum state tomography. We use this framework to examine how coherence and entanglement evolve during the execution of these algorithms and how their evolution relates to algorithmic performance. Specifically, we investigate these resources in two settings: combinatorial optimization, where we analyze their evolution as the algorithms approach optimal solutions, and quantum chemistry, where we examine their behavior in relation to electron correlation effects in molecular ground states.


The remainder of this article is organized as follows. Sec. \ref{sec:VQAs} introduces the variational quantum algorithms considered in this work, namely QAOA, FALQON, FALQON+, and VQE. Sec. \ref{sec:method} presents the resource-quantification framework based on the Bures distance and Uhlmann's theorem, including its variational formulation and adaptation to the algorithms studied here. Sec. \ref{sec:results} presents and discusses the results, analyzing the evolution of quantum coherence and entanglement in QAOA, FALQON, and FALQON+ for the Minimum Vertex Cover problem, as well as in VQE for the hydrogen molecule. Finally, Sec. \ref{sec:conclusion} summarizes our main findings and discusses possible directions for future research.


\section{Quantum optimization algorithms}
\label{sec:VQAs}


In this section, we review the quantum algorithms investigated in this work: QAOA, FALQON, FALQON+, and VQE. We first introduce the basic principles of VQAs and then describe the operating principles and mathematical formulations of each algorithm, establishing the foundation for the comparative analysis of their performance and the evolution of quantum coherence and entanglement presented in the following sections.

\subsection{Variational Quantum Algorithms}

In the NISQ era, VQAs have emerged as a prominent computational paradigm due to their flexibility in adapting to the limitations of contemporary quantum hardware, including finite coherence times, noise, limited qubit counts, and the lack of fault-tolerant quantum error correction. VQAs employ a hybrid quantum-classical architecture in which a classical optimizer iteratively updates a set of variational parameters, denoted by \(\pmb{\theta}\), to minimize a problem-dependent cost function. Formally, the optimization task is to determine
\begin{equation}
\pmb{\theta}^{*} = \arg\min_{\pmb{\theta}} C(\pmb{\theta}),
\end{equation}
where \(\pmb{\theta}^{*}\) denotes an optimal parameter vector and \(C(\pmb{\theta})\) is the cost function. In a general formulation, the cost function can be expressed as
\begin{equation}
C(\pmb{\theta}) = \sum_k f_k\left(\mathrm{Tr}\left[O_k U(\pmb{\theta}) \rho_k U^{\dagger}(\pmb{\theta})\right]\right),
\end{equation}
where \(\{f_k\}\) is a set of real-valued functions, \(\{O_k\}\) is a set of observables, \(\{\rho_k\}\) is a set of initial quantum states, and \(U(\pmb{\theta})\) is a parameterized unitary transformation. The latter is typically implemented as a sequence of circuit blocks
\begin{equation}
U(\pmb{\theta}) = \prod_{l=1}^{L} U_l(\pmb{\theta}_l),
\end{equation}
where \(L\) is the number of circuit blocks, and each \(U_l(\pmb{\theta}_l)\) represents a sequence of quantum gates that may depend on a subset \(\pmb{\theta}_l\) of the variational parameters. The choice of circuit architecture is a key determinant of VQA performance, as it influences the expressivity of the accessible quantum states, the trainability of the circuit, and its suitability for the problem at hand \cite{Friedrich2023,Holmes2022,Friedrich2024,Hubregtsen2021}.


Despite their potential, VQAs face several challenges, including the occurrence of \textit{barren plateaus}, in which cost-function gradients vanish exponentially with system size, hindering the optimization of circuit parameters \cite{Holmes2022,Larocca2025,McClean2018,Wang2021,Friedrich2022,Cerezo2021_BP,Arrasmith2021}. These challenges highlight the importance of developing circuit architectures and optimization strategies that improve trainability while accounting for hardware constraints. Nevertheless, VQAs remain an active research direction for near-term quantum computing, owing to their hybrid architecture and flexibility.


A central aspect of VQAs is the optimization of the variational parameters \(\pmb{\theta}\). Many optimization strategies rely on gradient-based methods, such as gradient descent, which updates the parameters according to the following rule:
\begin{equation}
\pmb{\theta}^{(t+1)} = \pmb{\theta}^{(t)} - \eta \nabla_{\pmb{\theta}} C\left(\pmb{\theta}^{(t)}\right),
\end{equation}
where \(\pmb{\theta}^{(t)}\) denotes the parameter vector at iteration \(t\), \(\eta\) is the learning rate, and \(\nabla_{\pmb{\theta}} C(\pmb{\theta}^{(t)})\) is the gradient of the cost function evaluated at the current parameter vector.


In classical simulations, gradients can be computed using techniques such as \textit{backpropagation}, which are widely used in machine learning. On quantum hardware, however, gradient estimation generally requires evaluating the cost function through repeated circuit executions. A widely adopted approach is the \textit{parameter-shift rule} \cite{Vidal2018}, which estimates derivatives from cost-function evaluations at appropriately shifted parameter values. Although this method provides an effective means of estimating gradients, it can require a substantial number of circuit evaluations, particularly as the number of variational parameters increases. Reducing this measurement and computational overhead remains an important challenge in the development of efficient VQAs.


\subsection{Quantum Approximate Optimization Algorithm}
\label{app:QAOA}

Given a cost Hamiltonian \(H_C\) that encodes the objective function of an optimization problem, the Quantum Approximate Optimization Algorithm (QAOA) employs a hybrid quantum-classical framework to search for approximate optimal solutions. Originally introduced for combinatorial optimization, QAOA can be viewed as a discretized, variational adaptation of the Quantum Adiabatic Algorithm (QAA) \cite{Farhi2014,Albash2018}.

QAOA is based on two Hamiltonians: the cost Hamiltonian \(H_C\), which encodes the optimization objective, and a mixer Hamiltonian \(H_M\), which drives transitions between computational basis states. A standard choice for the mixer is the sum of Pauli-\(X\) operators acting on all qubits
\begin{equation}
    H_M = \sum_{j \in V} X_j,
\end{equation}
where \(X_j\) denotes the Pauli-\(X\) operator acting on the \(j\)-th qubit. The algorithm proceeds through the following steps \cite{Blekos2024}:

\begin{enumerate}
    \item \textbf{Initialization:} The quantum register is initialized in an equal superposition of all computational basis states by applying a Hadamard gate to each qubit. The resulting state is
    \begin{equation}
        |s\rangle = |+\rangle^{\otimes n}
        = \frac{1}{\sqrt{2^n}} \sum_{x \in \{0,1\}^n} |x\rangle.
    \end{equation}
    This state is an eigenstate of \(H_M\) corresponding to its highest eigenvalue.

    \item \textbf{Ansatz construction:} The parameterized quantum circuit, or \textit{ansatz}, is constructed by applying \(p\) alternating layers of cost and mixer unitaries. For the \(k\)-th layer, these unitaries are defined as
    \begin{equation}
        U_C(\gamma_k) = e^{-i\gamma_k H_C},
        \qquad
        U_M(\beta_k) = e^{-i\beta_k H_M},
    \end{equation}
    where \(\pmb{\gamma} = (\gamma_1,\ldots,\gamma_p)\) and \(\pmb{\beta} = (\beta_1,\ldots,\beta_p)\) are variational parameters. The resulting depth-\(p\) state is given by
    \begin{equation}
        |\psi(\pmb{\gamma},\pmb{\beta})\rangle
        = \prod_{k=1}^{p} U_M(\beta_k) U_C(\gamma_k) |s\rangle.
        \label{eq:QAOA_state}
    \end{equation}
    Here, the product is understood to act from right to left, so the cost unitary is applied before the mixer unitary within each layer.

    \item \textbf{Measurement and classical optimization:} The quantum circuit is executed for the current parameter values, and measurements in the computational basis are used to estimate the expectation value of the cost Hamiltonian:
    \begin{equation}
        C(\pmb{\gamma},\pmb{\beta})
        = \langle\psi(\pmb{\gamma},\pmb{\beta})|
        H_C
        |\psi(\pmb{\gamma},\pmb{\beta})\rangle.
        \label{eq:QAOA_cost}
    \end{equation}
    Unlike the QAA, which follows a prescribed time-dependent evolution, QAOA treats \(\pmb{\gamma}\) and \(\pmb{\beta}\) as variational parameters. A classical optimizer updates these parameters iteratively to minimize the cost function, aiming to identify high-quality solutions to the optimization problem.
\end{enumerate}




As the classical optimizer converges, the variational state is expected to approach a state with a lower expectation value of \(H_C\), thereby increasing the probability of measuring bitstrings \(x^*\) that encode optimal or near-optimal solutions to the problem.

\textit{Application to the Minimum Vertex Cover (MVC) problem \cite{Faro2026}:} To formulate the MVC problem within the QAOA framework, we map the binary variables in the corresponding Quadratic Unconstrained Binary Optimization (QUBO) formulation to Pauli-\(Z\) operators, denoted by \(Z_v\). Following the standard QAOA convention, we use the mapping
\begin{equation}
x_v = \frac{1-Z_v}{2}.
\end{equation}
Under this correspondence, the eigenvalue \(Z_v=-1\) represents a vertex included in the cover (\(x_v=1\)), whereas \(Z_v=+1\) represents a vertex excluded from the cover (\(x_v=0\)). This substitution transforms the classical objective function into a diagonal quantum cost Hamiltonian, whose expectation value is minimized during the QAOA optimization.

Applying this mapping to the QUBO formulation, omitting additive constant energy offsets proportional to the identity operator, 
we obtain the following cost Hamiltonian for the unconstrained QAOA implementation:
\begin{equation}
    H_C =
    \frac{3}{4}\sum_{(i,j)\in E}
    \left(Z_i\otimes Z_j+Z_i+Z_j\right)
    -\sum_{i\in V}Z_i,
\end{equation}
where \(Z_i\) denotes the Pauli-\(Z\) operator acting on the qubit associated with vertex \(i\). This expression follows the Hamiltonian structure implemented by the PennyLane library \cite{Bergholm2018} in the simulations reported here. Further details on the MVC formulation and its mapping to a cost Hamiltonian are provided in Appendix \ref{app:MVC}. Regarding that Hamiltonian, we observe that here we choose \(A=3\) and \(B=2\).

\subsection{Feedback-based Algorithm for Quantum Optimization}
\label{app:FALQON}



Unlike QAOA, which relies on a classical optimization loop to determine the variational circuit parameters, FALQON (Feedback-based Algorithm for Quantum Optimization) \cite{Magann2022} uses measurement-based feedback to update its control parameters without requiring a conventional iterative classical optimizer. Inspired by Quantum Lyapunov Control (QLC) \cite{Hou2012}, FALQON aims to reduce the computational overhead associated with variational parameter optimization as the circuit depth \(p\) increases.

FALQON is formulated in terms of the continuous-time dynamics of a quantum system governed by a drift Hamiltonian, identified with the cost Hamiltonian \(H_C\), and a control Hamiltonian, identified with the mixer Hamiltonian \(H_M\). The state \(|\psi(t)\rangle\) evolves according to the Schr\"odinger equation
\begin{equation}
    i\frac{d}{dt}|\psi(t)\rangle
    = \left(H_C+\beta(t)H_M\right)|\psi(t)\rangle,
\end{equation}
where \(\beta(t)\) is a time-dependent scalar control function.


The objective is to minimize the expectation value of the cost Hamiltonian 
$
\langle H_C\rangle
= \langle\psi(t)|H_C|\psi(t)\rangle.
$
To ensure that this expectation value is non-increasing over time, we evaluate its time derivative using the Schr\"odinger equation:
\begin{equation}
    \frac{d}{dt}\langle H_C\rangle
    = i\beta(t)\langle\psi(t)|[H_M,H_C]|\psi(t)\rangle
    = \beta(t)A(t),
\end{equation}
where
\begin{equation}
    A(t) \equiv i\langle\psi(t)|[H_M,H_C]|\psi(t)\rangle.
\end{equation}
The Quantum Lyapunov Control (QLC) approach seeks to enforce
\begin{equation}
    \frac{d}{dt}\langle H_C\rangle \leq 0,
    \qquad \forall\,t\geq 0.
\end{equation}
A sufficient choice is to define the feedback control \(\beta(t)\) such that \(\beta(t)A(t)\leq 0\). The simplest linear feedback law is
\begin{equation}
    \beta(t)=-A(t),
\end{equation}
which gives
\begin{equation}
    \frac{d}{dt}\langle H_C\rangle=-A(t)^2\leq 0.
\end{equation}
Thus, the expected cost is guaranteed to be non-increasing under this feedback law.


In the discrete-time, circuit-based implementation of FALQON, the continuous evolution is approximated using a first-order Trotter decomposition, yielding alternating unitary operations structurally similar to those in the QAOA ansatz. The cost-layer parameter is fixed to a small time step, \(\gamma_k=\Delta t\), while the mixer parameter \(\beta_k\) at layer \(k\) is determined by the feedback law evaluated using the state from the preceding layer:
\begin{equation}
    \beta_k = -A_{k-1}
    = -i\langle\psi_{k-1}|[H_M,H_C]|\psi_{k-1}\rangle.
\end{equation}
At each iteration, the quantum computer prepares the state \(|\psi_{k-1}\rangle\) and estimates the expectation value of the Hermitian observable \(i[H_M,H_C]\). This estimate is then used to determine \(\beta_k\), which sets the mixer evolution in the next circuit layer. In this way, FALQON uses measurement-based feedback to update the circuit parameters without a conventional outer-loop classical optimization procedure.



In addition, Ref. \cite{Magann2022-1} introduces a variant known as FALQON+. This approach uses the parameters generated by a low-depth FALQON execution to initialize the QAOA circuit. By providing an informed initial parameter set, FALQON+ aims to mitigate the challenges associated with random initialization in standard QAOA and potentially reduce the classical optimization overhead while improving the approximation ratio and the probability of sampling optimal solutions.

To initialize a QAOA circuit of depth \(p\) using the output of FALQON, the parameters are mapped as follows:
\begin{align}
    \beta_k^{\text{(QAOA)}}
    &= \beta_k^{\text{(FALQON)}}\Delta t,
    \quad \text{and} \\
    \gamma_k^{\text{(QAOA)}}
    &= \Delta t,
    \quad \forall k\in\{1,\ldots,p\}.
\end{align}
Here, the FALQON mixer-control values are rescaled by the time step \(\Delta t\) to provide initial values for the QAOA mixer parameters, while the fixed FALQON time step initializes all QAOA cost parameters. The resulting parameter set, \(\{\beta_k^{\text{(QAOA)}},\gamma_k^{\text{(QAOA)}}\}_{k=1}^{p}\), is then refined using a classical optimization algorithm.

\subsection{Variational Quantum Eigensolver for Quantum Chemistry}
\label{app:VQE}




The VQE \cite{Peruzzo2014,Tilly2022} is a hybrid quantum-classical algorithm designed to approximate the ground state and ground state energy of a molecular Hamiltonian \(H_m\). In molecular simulations, the interatomic distance \(R\) is a key parameter that determines the molecular geometry and the corresponding Hamiltonian \(H_m(R)\).

In this work, we investigate the hydrogen molecule (\(H_2\)) using the minimal STO-3G basis set. The fermionic molecular Hamiltonian is mapped to a qubit Hamiltonian using the Jordan-Wigner transformation, yielding a four-qubit representation of the electronic problem in this basis. We employ a physically motivated VQE ansatz constructed from particle-number-preserving Givens rotations, implemented through parameterized single and double excitations in a unitary coupled-cluster-inspired (UCCSD-like) circuit.

The algorithm is initialized in the Hartree-Fock (HF) state \cite{Rossi2025}, which provides a single-reference approximation to the molecular ground state. A parameterized unitary \(U(\pmb{\theta})\) is then applied to generate a trial state,
\begin{equation}
|\psi(\pmb{\theta})\rangle
=U(\pmb{\theta})|\psi_{\mathrm{HF}}\rangle,
\end{equation}
whose energy expectation value is minimized with respect to the variational parameters:
\begin{equation}
E(R)=\min_{\pmb{\theta}}
\langle\psi(\pmb{\theta})|
H_m(R)
|\psi(\pmb{\theta})\rangle.
\end{equation}
By extracting the optimized VQE state at different interatomic distances, we investigate how quantum coherence and entanglement evolve as the molecular bond stretches from the equilibrium geometry toward the strongly correlated regime.

\section{Method}
\label{sec:method}

In this section, we briefly review the quantum resource quantification method employed in this work. The central idea is to formulate resource quantification as an optimization problem that can be addressed using a variational quantum algorithm (VQA). A detailed presentation of the method and its theoretical foundations is provided in Ref. \cite{Friedrich2025}.

Let \(\rho\) denote the quantum state whose resource content is to be quantified, and let \(\sigma\) be a state that is free of the resource under consideration. The corresponding Bures-based resource measure is defined as
\begin{equation}
    \mathcal{R}_B(\rho)
    = \min_{\sigma,\,|\Phi(\sigma)\rangle}
    2\left(
    1-\sqrt{F\bigl(|\Psi(\rho)\rangle,|\Phi(\sigma)\rangle\bigr)}
    \right),
\end{equation}
where \(|\Psi(\rho)\rangle\) and \(|\Phi(\sigma)\rangle\) are purifications of \(\rho\) and \(\sigma\), respectively, and \(F\bigl(|\Psi(\rho)\rangle,|\Phi(\sigma)\rangle\bigr)=|\langle\Psi(\rho)|\Phi(\sigma)\rangle|^2\) denotes the quantum fidelity between the two purifications. Thus, for a given target state \(\rho\), resource quantification amounts to finding a resource-free state \(\sigma\) and a corresponding purification that minimizes the Bures-distance-based objective.



Because the fidelity is evaluated between pure states, it can be estimated directly using a SWAP test \cite{Buhrman2001}, providing a practical means of implementing the resource-quantification procedure within a variational quantum algorithm. This approach avoids the need for full quantum state tomography, which can become experimentally demanding as the system size increases and may hinder the practical implementation of conventional resource-quantification methods.

The purification \(|\Psi(\rho)\rangle\) is determined entirely by the target state \(\rho\) and the chosen purification procedure. If a quantum circuit that prepares this purification is known, it can be used directly. Otherwise, general-purpose purification techniques, such as the method proposed in Ref. \cite{Pinto2024}, can be employed to construct a suitable purification.



The purification \(|\Phi(\sigma)\rangle\), however, requires additional considerations. First, because it is determined by the optimization procedure, its explicit form is not known a priori. Second, it must correspond to a purification of a state \(\sigma\) that is free of the resource being quantified. Consequently, the parametrization of \(|\Phi(\sigma)\rangle\) must be tailored to the resource under consideration.

In the following, we describe the parametrizations adopted for \(|\Phi(\sigma)\rangle\) in the quantification of quantum entanglement and coherence.



\textbf{Entanglement quantification.} When entanglement is the resource of interest, the free states are the mixed separable states. For a bipartite system \(AB\), these states can be expressed as convex combinations of product states:
\begin{equation}
    \sigma_{\mathrm{sep}}^{AB}
    = \sum_{j=0}^{N-1} p_j
    |\psi_j\rangle\langle\psi_j|_A
    \otimes
    |\phi_j\rangle\langle\phi_j|_B,
\end{equation}
where \(p_j\geq 0\), \(\sum_{j=0}^{N-1}p_j=1\), and \(N\) denotes the number of terms in the decomposition, the so called cardinality.

A purification compatible with this set of states can be constructed as
\begin{equation}
\begin{aligned}
|\Phi(\sigma_{\mathrm{sep}}^{AB})\rangle
={}& U_C
C_{U_j^A}^{C_j\rightarrow A}
C_{U_j^B}^{C_j\rightarrow B}
V_C
|000\rangle_{ABC},
\end{aligned}
\end{equation}
where \(U_C\), \(U_j^A\), \(U_j^B\), and \(V_C\) are unitary operators that determine the purification circuit. Further details on the construction and derivation of this purification are provided in Ref. \cite{Friedrich2025}.


\textbf{Quantum coherence quantification.} When quantum coherence is the resource of interest, it is quantified with respect to a fixed reference basis \(\{|j\rangle\}\). The free states are the incoherent states, which are diagonal in this basis and can be expressed as
\[
    \sigma_{\iota}^{A}
    = \sum_j p_j |j\rangle_A\langle j|,
\]
where \(p_j\geq 0\) and \(\sum_j p_j=1\). A corresponding purification is given by
\begin{equation}
    |\Phi(\sigma_{\iota}^{A})\rangle
    = U_B C_{X(j)}^{A_j\rightarrow B}
    V_A |0\rangle_A\otimes|0\rangle_B,
\end{equation}
where \(U_B\) and \(V_A\) are arbitrary unitary operators, and \(C_{X(j)}^{A_j\rightarrow B}\) denotes the controlled operation specified in the purification circuit.


\textbf{Adaptation to VQAs.} Although the method proposed in Ref.~\cite{Friedrich2025} was originally developed for quantifying quantum resources in general quantum states, here we employ it to investigate the evolution of quantum resources during the execution of QAOA, FALQON, FALQON+, and VQE. In this framework, the target state \(\rho\) is the quantum state prepared by the corresponding variational circuit. For the combinatorial optimization algorithms QAOA, FALQON, and FALQON+, the target state is considered at each algorithmic depth \(p\). For example, in QAOA, it is given by \(|\psi_p(\pmb{\gamma},\pmb{\beta})\rangle\). For VQE applied to molecular hydrogen, the target state is instead considered as a function of the interatomic distance \(R\) and is represented by the optimized variational state \(|\psi(\pmb{\theta}^{*}(R),R)\rangle\), where \(\pmb{\theta}^{*}(R)\) denotes the optimized parameter vector at each distance. These states provide the basis for examining how quantum coherence and entanglement vary throughout the optimization process and with molecular geometry, respectively, and for investigating their relationship with algorithmic performance.

\section{Results}
\label{sec:results}




In this section, we present the results of our investigation into the relationship between algorithmic performance and the evolution of quantum coherence and entanglement. We consider two distinct application domains: combinatorial optimization, where QAOA, FALQON, and FALQON+ are applied to the Minimum Vertex Cover (MVC) problem defined in Appendix~\ref{app:MVC}, and quantum chemistry, where VQE is used to approximate the ground state of the hydrogen molecule (\(H_2\)).

\begin{figure}[b]
    \centering
    \includegraphics[width=0.58
    \linewidth]{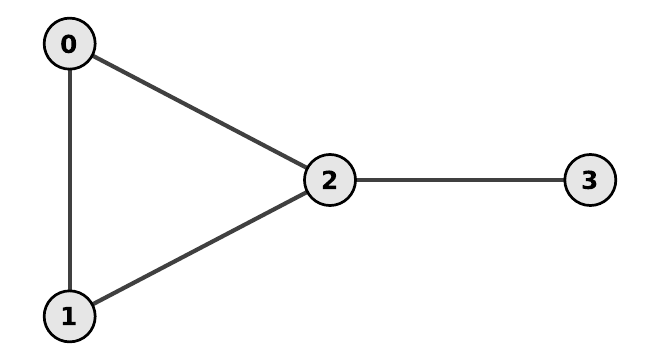}
    \caption{Topology of the graph used as the computational instance of the Minimum Vertex Cover (MVC) problem. The graph has four vertices ($|V|=4$) and four edges ($|E|=4$), forming a triangle connected to a pendant vertex.}
    \label{fig:graph}
\end{figure}

For combinatorial optimization, we consider the graph shown in Fig.~\ref{fig:graph} and quantify the Bures entanglement (\(E_b\)) and Bures coherence (\(C_b\)) of the output states as functions of the circuit depth \(p\). To monitor algorithmic convergence, we evaluate the mean squared error (MSE) between the probability distribution obtained from the final quantum state and the distribution associated with the optimal classical solution, represented by a bitstring. For VQE, we analyze the evolution of quantum coherence and entanglement as functions of the interatomic distance \(R\), while monitoring convergence through the molecular ground-state energy.

To perform this analysis, we follow the procedure below, using QAOA as an illustrative example:
\begin{enumerate}
    \item We fix the variational circuit depth \(p\), thereby defining the parameterized quantum state in Eq.~(\ref{eq:QAOA_state}).
    \item We optimize the variational parameters \(\pmb{\gamma}\) and \(\pmb{\beta}\) by minimizing the cost function \(C(\pmb{\gamma},\pmb{\beta})\) defined in Eq.~(\ref{eq:QAOA_cost}).
    \item Once the optimization converges, we store the final state \(|\psi(\pmb{\gamma}^{*},\pmb{\beta}^{*})\rangle\), corresponding to the optimized parameter values.
    \item We then use this state to obtain its computational-basis probability distribution and quantify its quantum resources using the variational resource-quantification method described in Sec.~\ref{sec:method}.
\end{enumerate}

An analogous procedure is followed for VQE. Instead of fixing the circuit depth \(p\), we fix the interatomic distance \(R\), optimize the variational parameters to minimize the molecular energy, and use the resulting optimized state to evaluate the corresponding Bures-based resource measures.

\subsection{QAOA}

\begin{figure}[ht]
    \centering
    \includegraphics[width=1\linewidth]{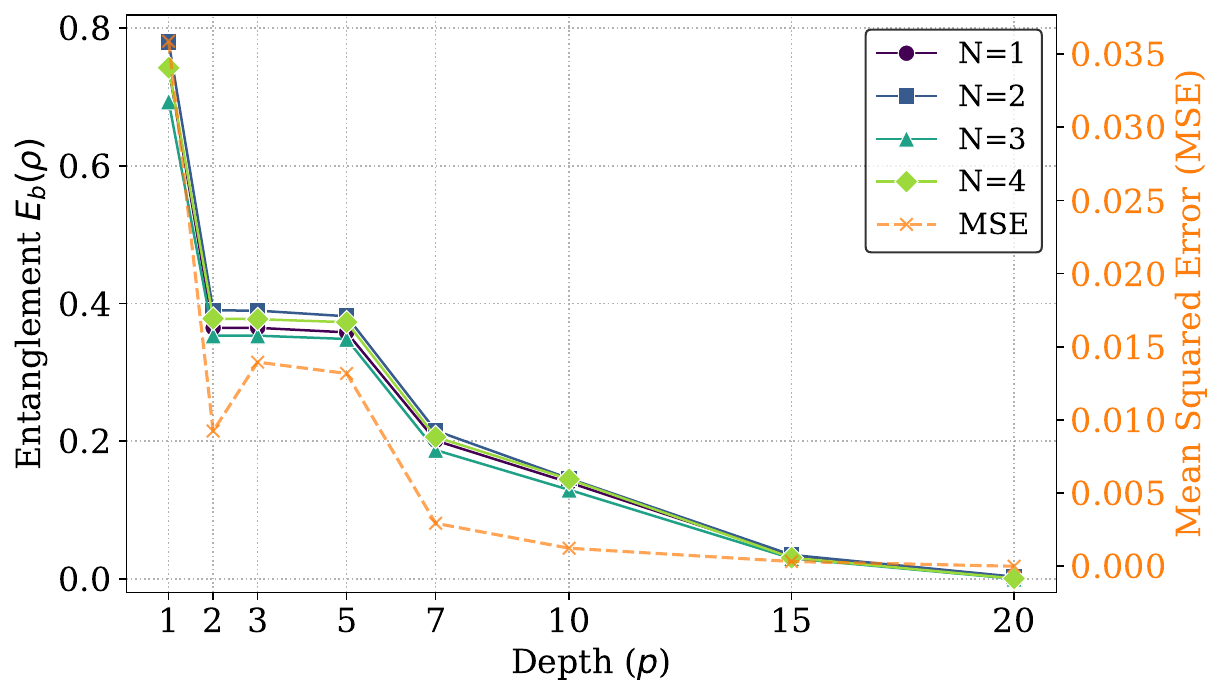}
    \caption{Evolution of the Bures entanglement $E_b$ and mean squared error (MSE) as functions of the QAOA circuit depth $p$. Entanglement initially increases, reaching a maximum at shallow depth, and subsequently decreases as the algorithm approaches convergence ($MSE \to 0$). The curves for $N=1,\ldots,4$ correspond to different cardinalities of the learner circuit's state. The hyperparameters used for all values of $N$ are $l_1=2$ (number of layers in the parametrization of $V_C$), $l_2=16$ (number of layers in the parametrization of $U_C$), $l=3$ (number of layers in the parametrizations of $U_A$ and $U_B$), learning rate $\eta=0.1$, and $100$ training epochs.}
    \label{fig:graph_qaoa_ent}
\end{figure}

Figure~\ref{fig:graph_qaoa_ent} shows the evolution of the Bures entanglement \(E_b\) and the mean squared error (MSE) as functions of the circuit depth \(p\) for the QAOA state \(|\psi(\pmb{\gamma},\pmb{\beta})\rangle\). As the circuit depth increases, the MSE generally decreases, indicating progressive convergence toward the optimal classical solution. Increasing the circuit depth expands the family of states accessible to the ansatz, potentially allowing the algorithm to identify better solutions.

As shown in Fig.~\ref{fig:graph_qaoa_ent}, the entanglement reaches its maximum at the first algorithmic layer (\(p=1\)), consistent with the nonlocal correlations generated by the initial QAOA layer. From \(p=2\) onward, the entanglement decreases and remains approximately constant until \(p=5\), following a trend similar to the MSE plateau over the same interval. At greater depths, the MSE resumes its decrease toward zero, while the entanglement continues to decay. This behavior suggests a relationship between the evolution of entanglement and the convergence dynamics of QAOA for the problem instance considered. The initial circuit layers generate quantum correlations, whereas the later stages approach a solution encoded in a computational-basis state, for which entanglement vanishes.

Finally, the entanglement quantification depends on the cardinality \(N\) used in the resource-quantification procedure, making its choice important for obtaining reliable estimates. For the present example, however, Fig.~\ref{fig:graph_qaoa_ent} indicates that varying \(N\) has only a minor effect on the reported results.

\begin{figure}[ht]
    \centering
    \includegraphics[width=1\linewidth]{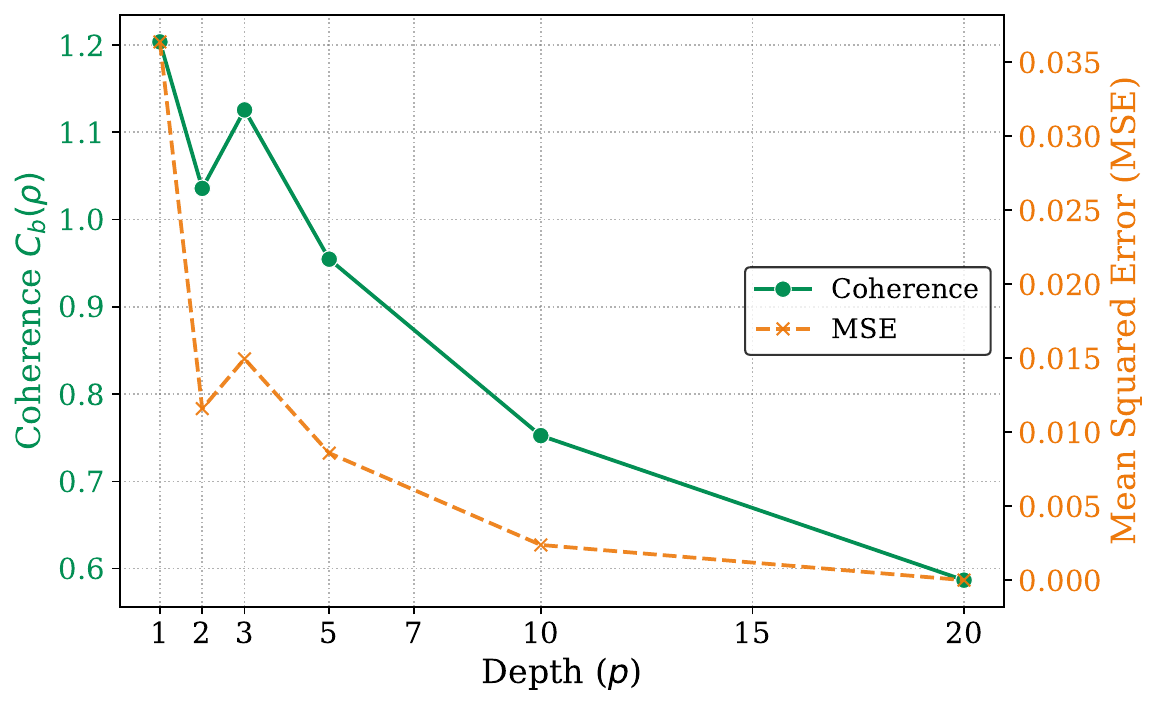}
    \caption{Evolution of the Bures coherence $C_b$ and mean squared error $MSE$ as functions of the QAOA circuit depth $p$. The decrease in coherence indicates a reduction in the state's coherence relative to the computational basis as the algorithm progresses toward a solution. The hyperparameters used in the resource-quantification procedure are $l_1=2$ (number of layers in the parametrization of $V_A$), $l_2=16$ (number of layers in the parametrization of $U_B$), learning rate $\eta=0.1$, and $100$ training epochs.}
    \label{fig:graph_qaoa_coherence}
\end{figure}

Figure~\ref{fig:graph_qaoa_coherence} shows the evolution of the Bures coherence \(C_b\) and the mean squared error (MSE) as functions of the QAOA circuit depth \(p\). Both metrics exhibit an overall decreasing trend as the circuit depth increases: the MSE decreases as the algorithm approaches an optimal solution, while the coherence also declines. This behavior suggests that the evolution of coherence is related to the convergence dynamics of QAOA for the problem considered.

A small oscillation occurs between \(p=2\) and \(p=3\), where both the MSE and coherence increase slightly. This indicates that adding a circuit layer does not necessarily improve the solution immediately, as the enlarged parameter space can make the optimization trajectory non-monotonic. At greater depths, both metrics resume their downward trends, and the MSE eventually approaches zero.

Unlike the case in which the algorithm converges to a unique optimal bitstring, the coherence remains finite even when the MSE becomes negligible. This behavior is consistent with the degeneracy of the present MVC instance, which has two optimal solutions, \(1010\) and \(0110\). If the final QAOA state is a coherent superposition of these optimal bitstrings, it can retain nonzero coherence in the computational basis even when the cost function has converged to its optimal value. The residual coherence may therefore reflect the structure of the degenerate solution space rather than a failure of the optimization procedure. This observation highlights a possible connection between the quantum resources of the output state and the properties of the underlying optimization problem.

\subsection{FALQON}



Although FALQON employs an alternating-layer circuit ansatz similar to that of QAOA, its feedback-based parameter assignment, inspired by Quantum Lyapunov Control (QLC), eliminates the need for a conventional outer-loop classical optimizer and leads to a different evolution of quantum resources.

\begin{figure}[b]
    \centering
    \includegraphics[width=1\linewidth]{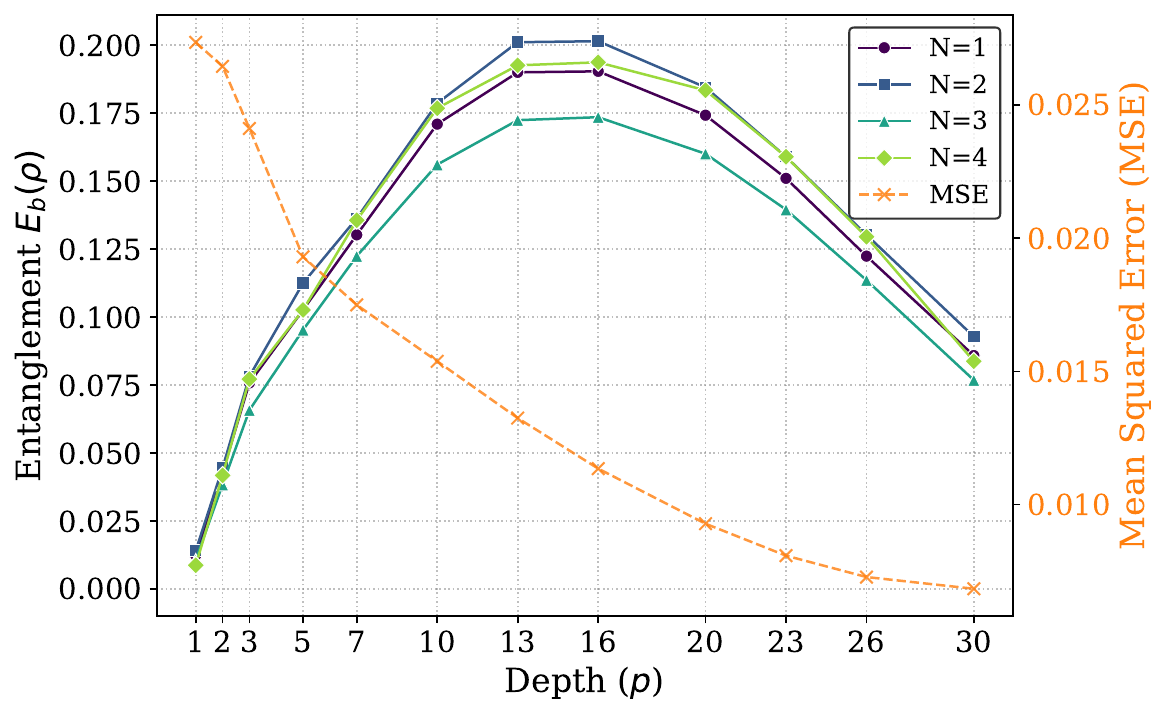}
    \caption{Evolution of $E_b$ and MSE across cricuit depth $p$ for the FALQON algorithm. Compared to QAOA, FALQON exhibits a much lower peak entanglement and a smoother resource trajectory. The general hyperparameters used for all values of $N$ are given by: $l_1=2$ (number of layers employed in the parametrization of $V_C$); $l_2=16$ (number of layers employed in the parametrization of $U_C$); $l=3$ (number of layers employed in the parametrization of $U_A$ and $U_B$); $\eta=0.1$ (learning rate); and 80 Epochs. }
    \label{fig:graph_falqon_ent}
\end{figure}

As shown in Fig.~\ref{fig:graph_falqon_ent}, the Bures entanglement \(E_b\) initially remains low and increases as the circuit depth grows, reaching a turning point at approximately \(p=16\). This increase indicates that the quantum correlations captured by the resource measure build up during the early stages of the algorithm. Beyond this point (\(p>16\)), the entanglement decreases as the MSE declines, suggesting a relationship between the evolution of entanglement and the algorithm's convergence toward an optimal solution. Finally, as in QAOA, the estimated Bures entanglement depends on the cardinality \(N\) used in the resource-quantification procedure. In the present analysis, \(N=3\) yields the lowest estimated Bures distance among the values considered.

\begin{figure}[t]
    \centering
    \includegraphics[width=1\linewidth]{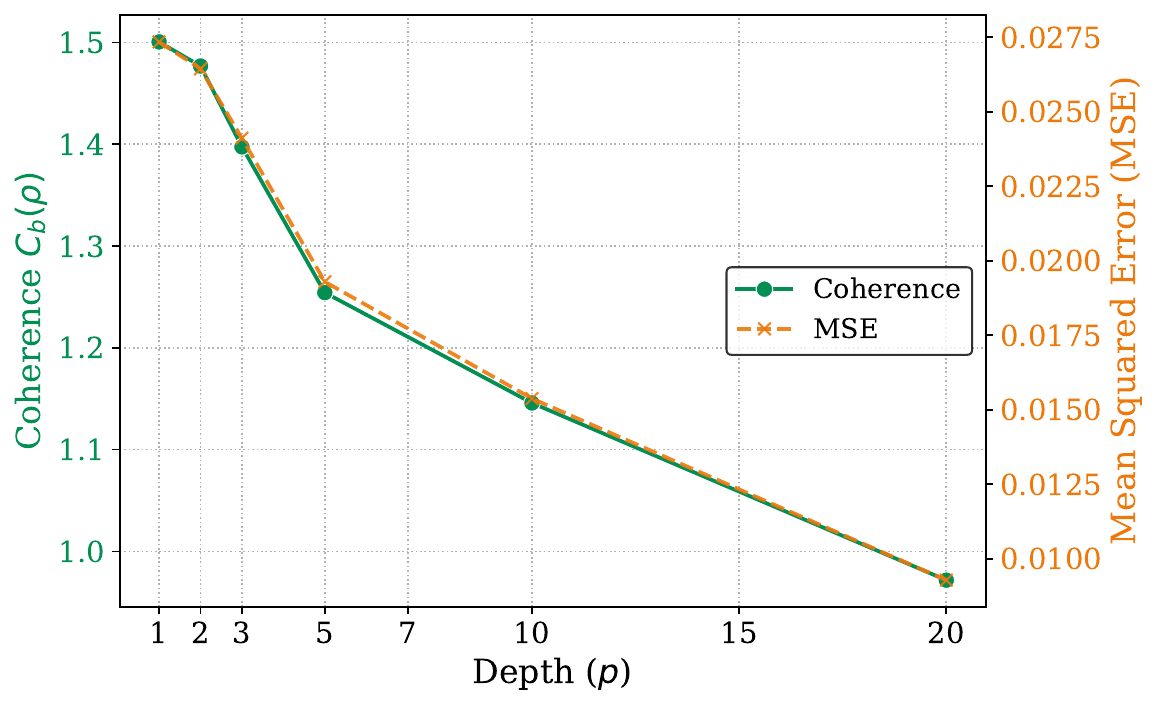}
    \caption{Evolution of $C_b$ and MSE for the FALQON algorithm. The coherence smoothly and monotonically decays in direct correlation with the algorithmic error. The hyperparameters used are given by: $l_1=4$ (number of layers employed in the parametrization of $V_A$); $l_2=12$ (number of layers employed in the parametrization of $U_B$); $\eta=0.1$; and 100 Epochs. }
    \label{fig:graph_falqon_coherence}
\end{figure}

Like QAOA, FALQON is initialized in the highly coherent uniform superposition \(|+\rangle^{\otimes n}\). As the circuit depth \(p\) increases, Fig.~\ref{fig:graph_falqon_coherence} shows a smooth, monotonic decrease in the Bures coherence \(C_b\). This trend closely follows the reduction in the MSE, suggesting a relationship between the loss of coherence and the algorithm's convergence toward an optimal solution. Since an optimal solution to the Minimum Vertex Cover problem can be represented by a computational-basis bitstring, the decreasing coherence is consistent with the evolution of the quantum state toward a solution with reduced coherence in the computational basis.

\subsection{FALQON+}



We also investigate the quantum resources associated with the hybrid FALQON+ approach, which uses feedback-derived parameters to initialize the classical optimization procedure. This initialization strategy leads to a resource-evolution profile that differs from those observed for standard QAOA and FALQON.

\begin{figure}[t]
    \centering
    \includegraphics[width=1\linewidth]{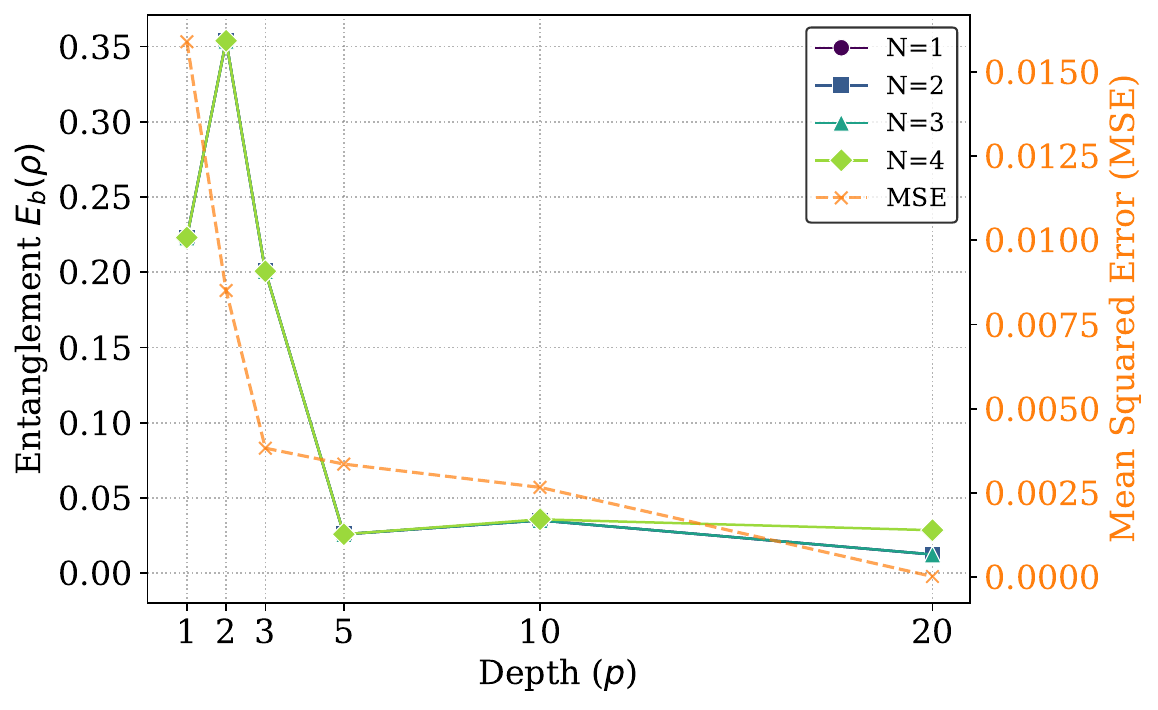}
    \caption{Evolution of Bures Entanglement ($E_b$) and Mean Squared Error (MSE) as a function of circuit depth $p$ for the FALQON+ algorithm. The general hyperparameters used for all values of $N$ are given by: $l_1=20$ (number of layers employed in the parametrization of $V_C$); $l_2=20$ (number of layers employed in the parametrization of $U_C$); $l=3$ (number of layers employed in the parametrization of $U_A$ and $U_B$); $\eta=0.1$; and $100$ Epochs. }
    \label{fig:graph_falqon_plus_ent}
\end{figure}

As shown in Fig.~\ref{fig:graph_falqon_plus_ent}, the Bures entanglement \(E_b\) exhibits a sharp peak at circuit depth \(p=2\), reaching a maximum value of approximately \(0.35\). This peak coincides with a pronounced initial decrease in the mean squared error (MSE). Beyond this point, the entanglement decreases rapidly, approaching zero for \(p\geq 5\). This behavior suggests that, for the problem considered, substantial entanglement is concentrated in the early stages of the optimization process, while the later stages exhibit little entanglement according to the measure employed. The simultaneous decrease in entanglement and MSE is consistent with the state approaching an optimal solution encoded in the computational basis, although the observed trends alone do not establish a causal relationship between entanglement and optimization performance. Finally, the results obtained for different cardinalities (\(N=1,\ldots,4\)) remain closely clustered, indicating that the estimated entanglement is relatively insensitive to \(N\) in this example.

\begin{figure}[ht]
    \centering
    \includegraphics[width=1\linewidth]{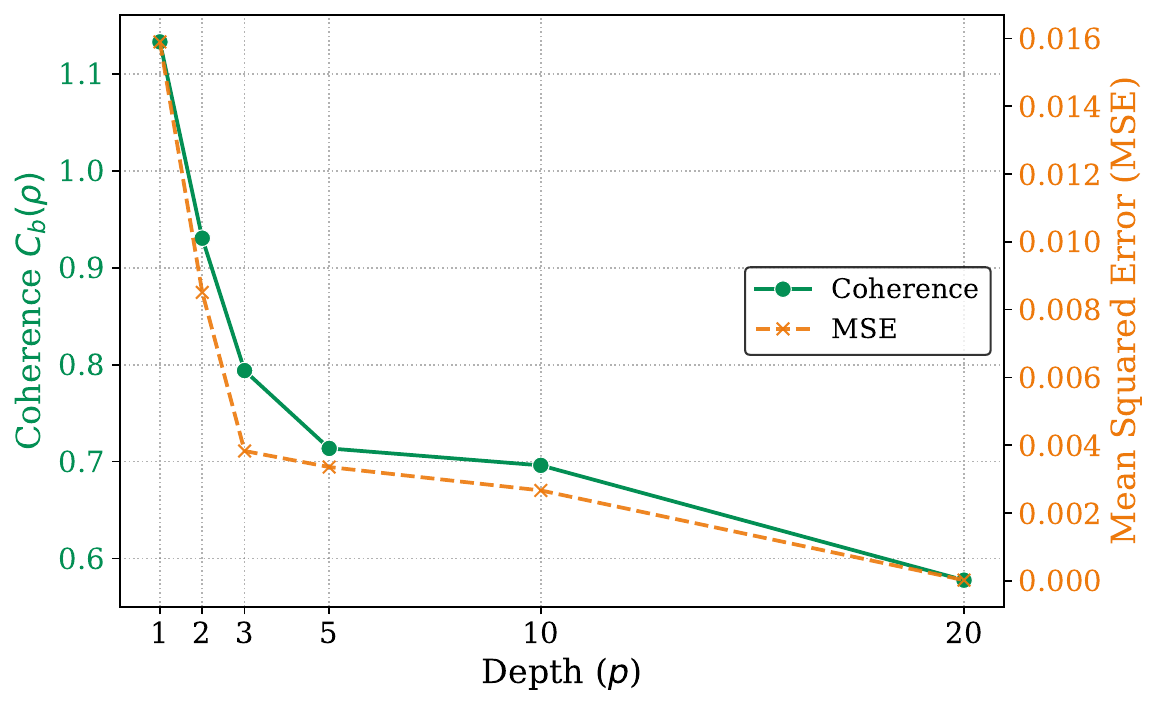}
    \caption{Evolution of Bures Coherence ($C_b$) and Mean Squared Error (MSE) as a function of circuit depth $p$ for the FALQON+ algorithm. The hyperparameters used are given by: $l_1=2$ (number of layers employed in the parametrization of $V_A$); $l_2=26$ (number of layers employed in the parametrization of $U_B$); $\eta=0.1$; and 100 Epochs.}
    \label{fig:graph_falqon_plus_coh}
\end{figure}

The evolution of quantum coherence provides further insight into the convergence behavior of FALQON+. Figure~\ref{fig:graph_falqon_plus_coh} shows that the Bures coherence \(C_b\) decreases smoothly and monotonically with circuit depth, following a trend similar to that observed for standard FALQON and accompanying the reduction in the optimization error. Starting from a highly coherent superposition at \(p=1\), the coherence progressively decreases as the circuit depth increases and the MSE approaches zero. This parallel evolution suggests a relationship between the reduction in coherence and the convergence of the algorithm toward the optimal solution subspace.

As observed for QAOA, the coherence does not vanish completely but approaches a finite residual value. This behavior is consistent with the degeneracy of the MVC instance, which has two optimal bitstrings, \(1010\) and \(0110\). If the final state retains a coherent superposition of these solutions, it can exhibit nonzero coherence in the computational basis even when the optimization error is negligible. Thus, the residual coherence may reflect the structure of the degenerate solution space rather than a failure of the optimization procedure.

\subsection{VQE}



Finally, we turn from combinatorial optimization to quantum chemistry and investigate the quantum resources associated with the VQE description of the hydrogen molecule (\(H_2\)) as a function of the interatomic distance \(R\), measured in Bohr radii (\(a_0\)).

\begin{figure}[h]
    \centering
    \includegraphics[width=1\linewidth]{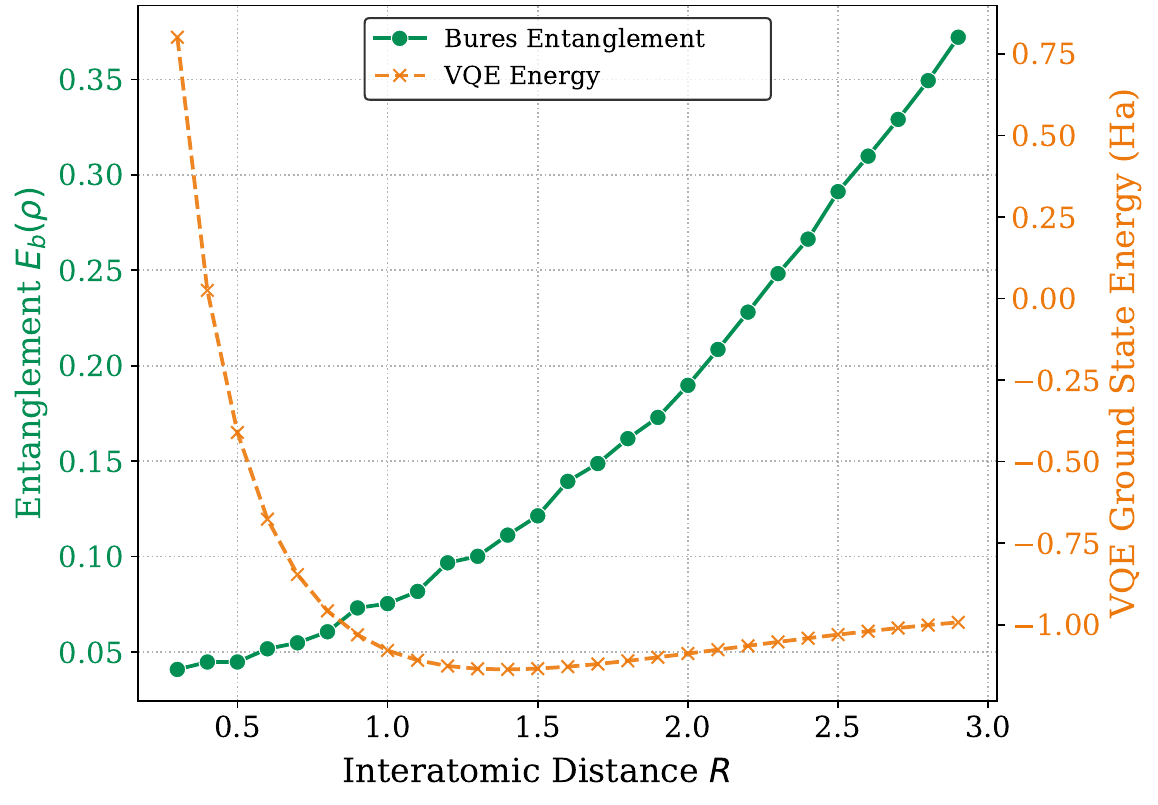}
    \caption{Bures entanglement ($E_b$) and VQE ground-state energy as functions of the interatomic distance (R) (in Bohr radii, $a_0$) for the $H_2$ molecule. The resource-quantification hyperparameters, for a fixed cardinality $N=4$, are $l_1=1$ (number of layers in the parametrization of $V_C$), $l_2=1$ (number of layers in the parametrization of $U_C$), $l=3$ (number of layers in the parametrizations of $U_A$ and $U_B$), learning rate $\eta=0.1$, and $100$ training epochs.}
    \label{fig:graph_vqe_ent}
\end{figure}

As shown in Fig.~\ref{fig:graph_vqe_ent}, the molecular energy follows the expected potential energy profile, reaching a minimum near the equilibrium bond length, \(R\approx 1.40\,a_0\). The Bures entanglement \(E_b\), evaluated using a fixed ancilla cardinality of \(N=4\), is relatively low near this equilibrium geometry. This behavior is consistent with the ground state being well approximated by a single-reference description in this regime. As the bond is stretched beyond \(R\approx 1.5\,a_0\), the entanglement increases, indicating a greater contribution from quantum correlations in the optimized VQE state. This trend is consistent with the onset of stronger static correlation as the covalent bond dissociates. In the dissociation regime, an accurate description of the electronic structure requires accounting for the changing contributions of multiple electronic configurations, which can be reflected in increased entanglement \cite{Huang2005, Ding2020}.

\begin{figure}[h]
    \centering
    \includegraphics[width=1\linewidth]{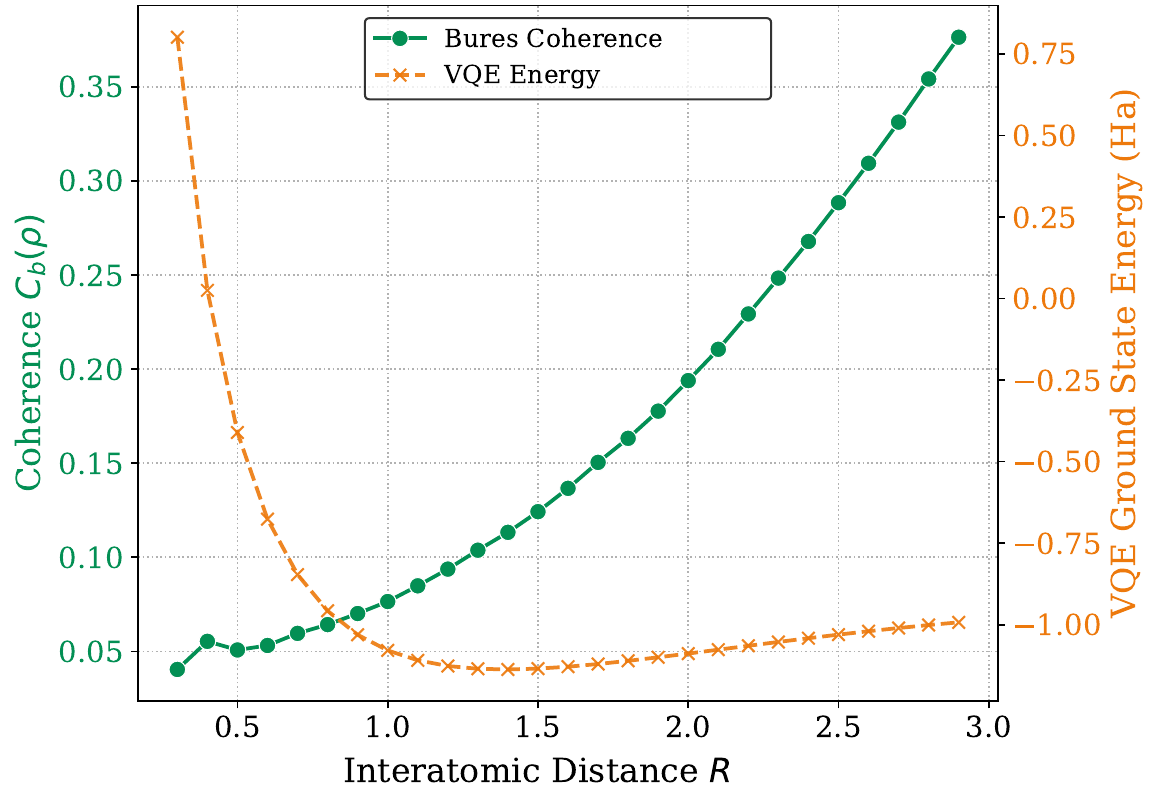}
    \caption{Evolution of Bures Coherence ($C_b$) and VQE Ground State Energy as a function of the interatomic distance $R$ (Bohr) for the $H_2$ molecule. The hyperparameters used are given by: $l_1=1$ (number of layers employed in the parametrization of $V_A$); $l_2=8$ (number of layers employed in the parametrization of $U_B$); $\eta=0.1$; and $100$ Epochs.}
    \label{fig:graph_vqe_coh}
\end{figure}

The evolution of quantum coherence, shown in Fig.~\ref{fig:graph_vqe_coh}, exhibits a trend similar to that observed for entanglement. The Bures coherence \(C_b\) remains relatively low near the equilibrium bond length (\(R\approx 1.40\,a_0\)), where the molecular ground state is well approximated by the Hartree-Fock reference configuration \(|1100\rangle\) in the Jordan-Wigner computational basis. As the bond stretches and the molecular energy increases toward the dissociation limit, the coherence increases. This behavior is consistent with the increasing contribution of electronic configurations beyond the Hartree-Fock reference, particularly the doubly excited configuration \(|0011\rangle\), which contributes to quantum superposition in the computational basis. Taken together, the entanglement and coherence results suggest that the resource-quantification framework captures changes in the electronic structure as the molecular bond is stretched.

\section{Conclusions}
\label{sec:conclusion}

In summary, we employed a variational framework to quantify and analyze two fundamental quantum resources--entanglement and coherence--during the execution of variational quantum algorithms (VQAs). By combining Uhlmann's theorem with the Bures metric and an auxiliary variational circuit, we estimated these resources in the output states of QAOA, FALQON, and FALQON+ applied to a minimum vertex cover (MVC) problem at different circuit depths, as well as in the VQE states describing the dissociation profile of the hydrogen molecule (\(H_2\)) at different interatomic distances.

For combinatorial optimization, our results reveal distinct patterns in the evolution of quantum resources across the three algorithms. In QAOA, the Bures entanglement initially increases, reaches a maximum at shallow circuit depth, and subsequently decreases as the algorithm approaches convergence. FALQON exhibits a different trajectory, with entanglement increasing up to an intermediate depth before declining, while its Bures coherence decreases monotonically with increasing depth. FALQON+ displays a pronounced entanglement peak at \(p=2\), followed by a rapid decline to nearly zero at greater depths, alongside a gradual reduction in coherence. These trends indicate that the evolution of entanglement and coherence is algorithm-dependent and can be correlated with changes in optimization performance. The coherence remaining at convergence is consistent with the possibility of a coherent superposition of degenerate optimal bitstrings, although degeneracy alone does not guaranty nonzero coherence.

The observed early-depth entanglement peak is also consistent with findings reported in related studies. Analyzes of QAOA using quantum Fisher information have identified substantial contributions from early entangling layers to parameter sensitivity and inter-parameter correlations, with additional layers producing diminishing or non-monotonic effects \cite{Sarmina2025}. Similarly, state-vector and matrix product state (MPS) simulations have reported intermediate-depth entanglement peaks followed by a decline in studies of classical combinatorial optimization problems \cite{Dupont2022, Nakhl2024}. These comparisons provide context for our results, while the precise resource trajectories remain dependent on the problem instance, algorithm, and quantification procedure.



Our results provide a complementary perspective on quantum resource quantification, consistent with tensor-network studies that have reported a decline in entanglement at greater circuit depths for certain combinatorial optimization problems. The variational Bures framework extends this analysis by jointly tracking entanglement and coherence throughout algorithm execution. However, such trends are not universal: while entanglement may decrease as an algorithm approaches a computational-basis solution, it can behave differently when the target state requires persistent quantum correlations. For example, studies of QAOA formulations for quantum relaxations of approximate optimization problems (QRAO) have reported monotonically increasing entanglement under certain conditions \cite{He2025}. These contrasting behaviors underscore the dependence of quantum-resource dynamics on the problem and algorithm considered.

Overall, the variational Bures framework provides a potentially useful diagnostic tool for characterizing quantum resources in near-term algorithms. In our combinatorial optimization results, the low entanglement and reduced coherence observed at greater circuit depths, as the algorithms approach convergence, suggest that the corresponding output states may admit more efficient classical representations. This observation is consistent with tensor-network simulations showing that restricting entanglement does not necessarily prevent QAOA from solving classical optimization problems, provided that sufficient circuit depth is available \cite{Sreedhar2022}. Nevertheless, low entanglement and coherence alone do not establish that a circuit can be executed reliably on noisy intermediate-scale quantum (NISQ) hardware: noise, gate errors, measurement overhead, and circuit depth remain important practical constraints. Further investigation is therefore needed to determine how resource estimates obtained with this framework can inform the implementation and evaluation of near-term quantum algorithms.

\appendix
\section{Minimum Vertex Cover and Ising Formulation}
\label{app:MVC}

The Minimum Vertex Cover (MVC) problem is a well-known NP-hard problem in graph theory. Given an undirected graph \(G=(V,E)\), the objective is to find a minimum-cardinality subset \(S\subseteq V\) such that every edge in \(E\) has at least one endpoint in \(S\).

To formulate the MVC problem as an Ising model, or equivalently as a Quadratic Unconstrained Binary Optimization (QUBO) problem \cite{Djidjev2018}, we introduce a binary variable \(y_i\in\{0,1\}\) for each vertex \(i\in V\). The variable \(y_i=1\) indicates that vertex \(i\) belongs to the cover, whereas \(y_i=0\) indicates that it does not. The cost function is constructed as the sum of a constraint-penalty term and an objective term,
\begin{equation}
H=H_A+H_B,
\end{equation}
where \(H_A\) penalizes violations of the vertex-cover constraints and \(H_B\) encodes the objective of minimizing the number of selected vertices \cite{Lucas2014}.

For a valid vertex cover, every edge \((i,j)\in E\) must have at least one endpoint included in the cover. Thus, an edge is uncovered only when \(y_i=y_j=0\). This constraint is enforced by assigning a positive penalty \(A>0\) to each uncovered edge:
\begin{equation}
    H_A=A\sum_{(i,j)\in E}(1-y_i)(1-y_j).
\end{equation}
This term vanishes when all edges are covered and is strictly positive whenever at least one edge is uncovered.

The objective is to minimize the number of vertices in the cover. We therefore assign a positive cost \(B>0\) to each selected vertex:
\begin{equation}
    H_B=B\sum_{i\in V}y_i.
\end{equation}
The penalty coefficient \(A\) must be sufficiently large relative to \(B\) to ensure that minimizing the total cost favors valid vertex covers over configurations that leave edges uncovered. In particular, choosing \(A>B>0\) is sufficient for this formulation \cite{Lucas2014}. The resulting QUBO model uses one binary variable per vertex and therefore requires \(|V|\) binary variables. Under the corresponding Ising mapping, this translates to \(|V|\) spin variables or qubits.


\end{document}